# The calculation of average kernel with Gauss-Laguerre quadrature for double integrals

Xie Mingliang

State Key Laboratory of Coal Combustion, Huazhong University of Science and Technology, Wuhan 430074, China
Corresponding Email: mlxie@mail.hust.edu.cn

**Abstract:**
The use of average kernel method based on the Laplace transformation can significantly simplify the procedure for obtaining approximate analytical solution of Smoluchowski equation. However, this method also has its own shortcomings, one of which is the computational complexity of the binary Laplace transformation for a nonlinear collision kernel. In this study, a universal algorithm based on the Gauss-Laguerre quadrature for treating the double integral is developed to obtain easily and quickly pre-exponential factor of the average kernel. Furthermore, the corresponding error estimate are also provided.

**Introduction**

The Smoluchowski equation is one of the basic equations in aerosol science and technology, and also one of the main equations of kinetic molecular theory, which has a wide range of applications in the field of mathematics and engineering. For a mono-variant problem, the classical Smoluchowski equation takes the form (Friedlander, 2000)

$$\frac{\partial n(v,t)}{\partial t} = \frac{1}{2}\int_0^v \beta(v_1, v-v_1)n(v_1)n(v-v_1)dv_1 - \int_0^\infty \beta(v_1,v)n(v_1)n(v)dv_1 \qquad (1)$$

in which $n(v,t)dv$ is the number density of particles per unit spatial volume with particle volume from $v$ to $v+dv$ at time $t$; and $\beta$ is the kernel of coagulation.

In the past 100 years, the Smoluchowski equation has been solved analytically only for a limited number of known simple collision kernels, such as constant, additive and multiplicative kernels (Leyvraz, 2003). In the real world, the collision kernel under different physical conditions usually takes a nonlinear form. The convolution in the Smoluchowski equation is usually impossible to integrate if the collision kernel is a function of the particle size. In 1940, Schumann proposed the concept of the average kernel to approximate the actual collision kernel, which significantly simplifies the analytical solution procedure of the Smoluchowski equation (Schumann, 1940).

By carefully analyzing Schumann's work, the average kernel method also has its own drawbacks. Firstly, the analytical solution is independent of the average kernel and the boundary condition (Pan et al., 2024). Secondly, there is an inherent contradiction between the analytical solution and the experimental data (Xie, 2024). Thirdly, the binary Laplace transformation for a nonlinear collision kernel has higher computational complexity (Makoveeva and Alexandrov, 2023).

Coincidentally, an iterative direct numerical simulation (iDNS) has been proposed recently (Xie and He, 2022). This method uses the asymptotic solution of Taylor-series expansion method of moments (TEMOM) or analytical solution of average kernel method (AK) method as the initial condition (Xie, 2023; Pan et al., 2024), and the self-preserving size distribution is achieved through an iterative algorithm. The corrected similarity solution is not only consistent with the experimental data but also consistent with the properties of the original kernel. Therefore, the first two defects of average kernel method have been effectively addressed and overcome, and an AK-iDNS framework is proposed to solve the Smoluchowski equation accurately and efficiently (Pan et al., 2024; Xie, 2024).

In the previous work, we treated the definition of average kernel as a Laplace transform (Pan et al., 2024; Xie, 2024), where the binary Laplace transform projects a spatial surface onto a point on the symmetry plane. Utilizing the homogeneity of kernel, the projection point is replaced by an intersection point between the surface and the symmetric plane. This treatment greatly simplifies the calculation of the average kernel and achieves the same scaling law as that of TEMOM. However, the simplified computational method of the average kernel may produce some deviations to some extent, and is inconsistent with the physical meaning defined by the average kernel. For example, the average kernel of sedimentation coagulation will be zeros, which is obviously unreasonable.

The purpose of this study is to address and improve the third drawback of the average kernel method, and to provide an effective numerical method for calculating the pre-exponential factor of the average kernel. This article will treat the definition of average kernel as the Gaussian-Laguerre quadrature for double integrals, and a compact and fast MATLAB code calculating the double Gauss-Laguerre integrals and its error are proposed. The content is roughly divided into four parts: part I introduces the average kernel method; part II constructs the expression and its truncation error of pre-exponential factor of the average kernel based on the Gauss-Laguerre quadrature for a double integral; part III establishes the universal algorithm of the Gauss-Laguerre quadrature in evaluating the abscissas and weights; the last part demonstrates the results of common collision kernels.

**The average kernel method**

To solve the Smoluchowski equation with collision kernel depending on the particle size, Schumann proposed the average kernel method (Schumann, 1940), which is defined with Laplace transformation as

$$\int_0^\infty \int_0^\infty \bar{\beta} \exp\left(-\frac{v_1+v}{u}\right) dv_1 dv = \int_0^\infty \int_0^v \beta(v_1, v) \exp\left(-\frac{v_1+v}{u}\right) dv_1 dv \qquad (2)$$

where $u$ is the algebraic mean volume of particle size distribution, and it is defined as $u = \int_0^\infty v n(v,t) dv / \int_0^\infty n(v,t) dv$. Through operation and reorganization, the average kernel can be represented as

$$\bar{\beta} = \frac{1}{u^2} \int_0^\infty \int_0^v \beta(v_1, v) \exp\left(-\frac{v_1+v}{u}\right) dv_1 dv \qquad (3)$$

For a homogeneous collision kernel, it has the following properties:

$$\begin{cases} \beta(\alpha v, \alpha v_1) = \alpha^q \beta(v, v_1) \\ \beta(v, v_1) = \beta(v_1, v) \\ \beta(v, v_1) \geq 0 \end{cases} \quad (4)$$

in which $q$ is power index, and $\alpha$ is the scale factor. It is easy to prove that the homogeneous collision kernels satisfy the following differential equation:

$$v \frac{\partial \beta}{\partial v} + v_1 \frac{\partial \beta}{\partial v_1} - q\beta = 0 \quad (5)$$

If the scale factor is $\alpha = 1/u$, the collision kernel can be expressed as

$$\beta(v, v_1) = u^q \beta\left(\frac{v}{u}, \frac{v_1}{u}\right) \quad (6)$$

Let the dimensionless particle volume be defined as

$$\eta = \frac{v}{u} \quad (7)$$

and the average kernel can be expressed as

$$\bar{\beta} = p u^q \quad (8)$$

where $p$ is a proportional factor, which can be calculated as

$$p = \int_0^\infty \int_0^{u\eta} e^{-\eta-\eta_1} \beta(\eta, \eta_1) d\eta_1 d\eta \quad (9)$$

which usually representing the total collision frequency of particle coagulating system. Due to the symmetry of the homogeneous collision kernel, $p$ can be simplified as

$$p = \frac{1}{2} \int_0^\infty \int_0^\infty e^{-\eta-\eta_1} \beta(\eta, \eta_1) d\eta_1 d\eta \quad (10)$$

According to the structure of the integrals, it is natural to think of approximating it with the Gauss-Laguerre quadrature. But it is a double integral, and the relevant numerical methods and error estimates have not been reported in the literature yet.

In addition, there is another way to get the population-averaged coagulation kernel, which is listed in Appendix I. Since the analytical solution of the Smoluchowski equation with averaged kernel is $\psi(\eta) = e^{-\eta}$, therefore, the average kernel based on Laplace transformation is equivalent to the population-averaged kernel.

**The Gauss-Laguerre quadrature and its remainder**

Consider the single integral $I$ with the $(n+1)$-point Gauss-Laguerre quadrature rule given by:

$$I = \int_0^\infty e^{-x} f(x) dx = \sum_{j=1}^{n+1} A_i f(x_i) + R_{n+1}(f) \quad (11)$$

where the abscissas $x_i$ are the zeros of the Laguerre polynomials $L_{n+1}(x)$, and $L_{n+1}(x)$ is defined as

$$L_{n+1}(x) = \frac{e^x}{(n+1)!} \frac{d^{n+1}(x^{n+1} e^{-x})}{dx^{n+1}} \quad (12)$$

and $A_i$ are the corresponding weights or Christoffel numbers, and $A_i$ can be expressed as

$$A_i = \int_{-1}^1 \prod_{i=0, i\neq k}^{n+1} \frac{x - x_i}{x_k - x_i} e^{-x} dx \quad (13)$$

On the assumption that the definition of $f$ may be continued into the complex $z$-plane, where $z = x + iy$. Then the remainder, or truncation error, of the quadrature rule is a contour integral.

The contour $C_z$ encloses the interval $-1 \leq Re(z) \leq 1$ and is such that the function $f$ is analytic on and within $C_z$. Then the remainder is given by (Donaldson and Elliott, 1972):

$$R_{n+1}(f) = \frac{1}{2\pi i} \oint k_{n+1}(z) f(z) dz \tag{14}$$

in which the function $k_{n+1}(z)$ is given by:

$$k_{n+1}(z) = \frac{\Pi_{n+1}(z)}{L_{n+1}(z)} \tag{15}$$

where

$$\Pi_{n+1}(z) = \int_0^\infty \frac{e^{-t} L_{n+1}(t)}{z-t} dt \tag{16}$$

Next, consider the double integral $II$ with $(n+1) \times (n+1)$-point Gauss-Laguerre quadrature rule given by:

$$II = \int_0^\infty \int_0^\infty e^{-x-x_1} f(x, x_1) dx_1 dx$$

$$= \sum_{i=1}^{n+1} \sum_{j=1}^{n+1} A_i A_j f(x_i, x_j) + R_{n+1,x_1} + R_{n+1,x} + O(R_{n+1,x}) \tag{17}$$

in which, the remainders are given by:

$$R_{n+1,x_1} = \frac{1}{2\pi i} \oint k_{n+1}(z) f(x, z) dz \tag{18a}$$

$$R_{n+1,x} = \frac{1}{2\pi i} \oint k_{n+1}(z) f(z, x_1) dz \tag{18b}$$

$$O(R_{n+1,x}) = -\frac{1}{2\pi i} \oint k_{n+1}(z) R_{n+1,x_1}(z) dz \tag{18c}$$

In the above, it is assumed that the remainder $R_{n+1,x_1}$ may be continued into the complex $z$-plane. In any case, the last term $O(R_{n+1,x})$ is essentially the remainder of remainder and henceforth it is assumed to be negligible and it is replaced by zero (Elliott et al., 2011). On combing these results, the double integral can be expressed as

$$II = \int_0^\infty \int_0^\infty e^{-x-x_1} f(x, x_1) dx_1 dx = Q_{n+1,n+1} + R_{n+1,n+1} \tag{19}$$

in which $Q_{n+1,n+1}$ and $R_{n+1,n+1}$ are given as follows:

$$Q_{n+1,n+1} = \sum_{i=1}^{n+1} \sum_{j=1}^{n+1} A_j A_j f(x_i, x_j) \tag{20}$$

and

$$R_{n+1,n+1} = 2 \int_0^\infty R_{n+1}(f(x,x)) dx + O(R_{n+1,x}) \tag{21}$$

**Evaluation points and weights of Gauss-Laguerre quadrature**

The Laguerre polynomials can be calculated recursively. Defining the first two polynomials as

$$L_0(x) = 1, \ L_1(x) = 1 - x; \tag{22}$$

And the recurrence relation of Laguerre polynomials for any $n \geq 1$ is,

$$L_{n+1}(x) = \frac{(2k+1-x) L_n(x) - n L_{n-1}(x)}{n+1}; \tag{23}$$

It can be written explicitly as

$$L_{n+1}(x) = \frac{1}{(n+1)!}[(-x)^{n+1} + (n+1)^2(-x)^n + \cdots$$
$$+(n+1)(n+1)!(-x) + (n+1)!] \tag{24}$$

And the abscissas $x_i$ can be obtained from the zeros of the Laguerre polynomials, i.e., $L_{n+1}(x) = 0$. The corresponding weights $A_i$ can be calculated explicitly as

$$A_i = \frac{x_i}{(n+1)^2[L_{n+1}(x_i)]^2} \tag{25}$$

Then the single integral $I$ and the double integral $II$ can be calculated.

For instance, the Laguerre polynomials $L_{n+1}(x)$ for $n = 10$ is

$$L_{11}(x) = \frac{x^{10}}{3628800} - \frac{x^9}{36288} + \frac{x^8}{896} - \frac{x^7}{42} +$$
$$\frac{7x^6}{24} - \frac{21x^5}{10} + \frac{35x^4}{4} - 20x^3 + \frac{45x^2}{2} - 10x + 1 \tag{26}$$

The corresponding zeros and weights for Gauss Laguerre quadrature can be found in Table 1.

The remainder for single integral is given by

$$R_{n+1}(f) = \int_0^\infty \frac{((n+1)!)^2}{(2n+2)!} f^{2n+2}(\xi) dx \tag{27}$$

The corresponding MATLAB code is listed in Appendix II.

Table 1. The zeros and weights of Gauss Laguerre quadrature for $n = 10$

| $i$ | $x_i$ | $A_i$ |
| --- | --- | --- |
| 1 | 0.1377 | 0.3084 |
| 2 | 0.7294 | 0.4011 |
| 3 | 1.8083 | 0.2180 |
| 4 | 3.4014 | 0.0620 |
| 5 | 5.5524 | 0.0095 |
| 6 | 8.3301 | 0.0007 |
| 7 | 11.8437 | 2.8E-05 |
| 8 | 16.2792 | 4.2E-07 |
| 9 | 21.9965 | 1.8E-09 |
| 10 | 29.9206 | 9.9E-13 |

**Results and Discussions**

In this section, several examples are used to demonstrate the accuracy of the Gauss-Laguerre quadrature for the double integral. Some common collision kernels for various mechanisms of particle interactions are listed bellows,

$$\beta_{SC} = \left(\eta_1^{\frac{1}{3}} + \eta^{\frac{1}{3}}\right)^3 \tag{28}$$

for shear coagulation (SC),

$$\beta_{SD} = \left(\eta^{\frac{1}{3}} + \eta_1^{\frac{1}{3}}\right)^3 \left|\eta^{\frac{1}{3}} - \eta_1^{\frac{1}{3}}\right| \tag{29}$$

for particles sedimentation under gravity (SD),

$$\beta_{FM} = (\eta_1^{-1} + \eta^{-1})^{\frac{1}{2}} \left(\eta_1^{\frac{1}{3}} + \eta^{\frac{1}{3}}\right)^2 \tag{30}$$

for Brownian coagulation in the free molecule regime (FM),

$$\beta_{CR} = \left(\eta_1^{-\frac{1}{3}} + \eta^{-\frac{1}{3}}\right)\left(\eta_1^{\frac{1}{3}} + \eta^{\frac{1}{3}}\right) \tag{31}$$

for Brownian coagulation in the continuum regime (CR), respectively. For brevity, some physical constants have been ignored in the expression of collision kernel.

The change of the value $Q_{n+1,n+1}$ and its error $\epsilon_n$ as the interpolation points increases is shown in Fig.1. and the error $\epsilon_n$ is defined as

$$\epsilon_n = |Q_{n+1,n+1} - Q_{n,n}| \tag{32}$$

As the number of evaluation points increases, the value of $Q_{n+1,n+1}$ gradually tends towards a fixed value; and the error decreases sharply, there appear to a constant slope in the error plot on the logarithmic coordinate system. According to the two-point formula of the line, it means that the following equation holds

$$\frac{\ln \epsilon_m - \ln \epsilon_n}{\ln m - \ln n} = C, \quad (m, n \in N) \tag{33}$$

in which $C$ is the constant slope. And it can be reorganized into

$$\frac{\epsilon_m}{\epsilon_n} = \left(\frac{m}{n}\right)^C \tag{34}$$

It can be extended to the real number field as

$$\epsilon_x = \epsilon_n \left(\frac{x}{n}\right)^C, \quad (x \in R) \tag{35}$$

And the truncation error can be calculated as

$$R_{n+1,n+1} = \int_{n+1}^{\infty} \epsilon_x dx = -\epsilon_n \left(\frac{n+1}{n}\right)^C \left(\frac{n+1}{C+1}\right); \quad (C < -1, C \in R) \tag{36}$$

In this article, the maximum number of the interpolation points is $n = 360$, and the corresponding error $\epsilon_n$, the constant slope $C$, $Q_{n+1,n+1}$ and its remainder $R_{n+1,n+1}$ are listed in Table 2.

Table 2. The numerical values related to the Gauss-Laguerre quadrature for double integrals

| Type | $\epsilon_n$ | $C$ | $Q_{n+1,n+1}$ | $R_{n+1,n+1}$ | $II$ |
|---|---|---|---|---|---|
| FM | $1.2344E - 4$ | $-1.5209$ | $6.9032$ | $0.0852$ | $6.9032 \pm 0.0852$ |
| CR | $2.5865E - 5$ | $-1.6572$ | $4.4025$ | $0.0141$ | $4.4025 \pm 0.0141$ |
| SC | $9.3561E - 7$ | $-2.3733$ | $6.8371$ | $0.0002$ | $6.8371 \pm 0.0002$ |
| SD | $7.4036E - 6$ | $-1.9750$ | $2.5861$ | $0.0027$ | $2.5861 \pm 0.0027$ |

The value of the pre-exponential factor $p$ of the average kernel is half of the double integrals $II$, and the average kernel values for different collision kenrels are listed in Table 3.

Table 3. The converged average kernels for a few common collision kernels

| Type | $\beta(v, v_1)$ | $\bar{\beta} = pu^q$ |
|------|-----------------|----------------------|
| FM | $\beta_{FM} = (v_1^{-1} + v^{-1})^{\frac{1}{2}}\left(v_1^{\frac{1}{3}} + v^{\frac{1}{3}}\right)^2$ | $\bar{\beta}_{FM} = 3.4516 u^{\frac{1}{6}}$ |
| CR | $\beta_{CR} = \left(v_1^{-\frac{1}{3}} + v^{-\frac{1}{3}}\right)\left(v_1^{\frac{1}{3}} + v^{\frac{1}{3}}\right)$ | $\bar{\beta}_{CR} = 2.2013$ |
| SC | $\beta_{SC} = \left(v_1^{\frac{1}{3}} + v^{\frac{1}{3}}\right)^3$ | $\bar{\beta}_{SC} = 3.4186 u$ |
| SD | $\beta_{SD} = \left(v_1^{\frac{1}{3}} + v^{\frac{1}{3}}\right)^3 \left|v_1^{\frac{1}{3}} - v^{\frac{1}{3}}\right|$ | $\bar{\beta}_{SD} = 1.2931 u^{\frac{4}{3}}$ |

**Conclusions**

In this study, a compact and fast MATLAB code based on the Gauss-Laguerre quadrature for the double integrals is proposed to calculate the value of the average kernel after proper normalization, which reduces the complexity of the traditional average kernel method in the literatures. The numerical method together with the corresponding code will lay the foundation for the promotion and application of the AK-iDNS framework when solving the Smoluchowski equation.

**Acknowledgement**

This research was funded by the National Natural Science Foundation of China with grant number 11972169.

**Appendix I: Population average kernel**

In the literature, the population-averaged coagulation kernel is defined as

$$\int_0^\infty \int_0^\infty \bar{\beta}\, n(v_1)n(v)\, dv_1 dv = \int_0^\infty \int_0^\infty \beta(v_1,v)\, n(v_1)n(v)\, dv_1 dv \tag{AI.1}$$

It can be calculated as

$$\bar{\beta} = \frac{\int_0^\infty \int_0^\infty \beta(v_1,v)\, n(v_1)n(v)\, dv_1 dv}{\int_0^\infty \int_0^\infty n(v_1)n(v)\, dv_1 dv} \tag{AI.2}$$

Based on the analytical similarity solution (Schumann, 1940; Pan et al., 2024),

$$\psi(\eta) = e^{-\eta} \tag{AI.3}$$

The particle number density can be expressed as

$$n(v,t) = \frac{N}{u}\exp\left[-\frac{v}{u}\right] \tag{AI.4}$$

where $N = \int_0^\infty n(v,t)dv$ is the total particle number. And the population-averaged kernel can be expressed as

$$\bar{\beta} = \frac{1}{u^2}\int_0^\infty \int_0^\infty \beta(v_1,v)\exp\left(-\frac{v_1+v}{u}\right) dv_1 dv \tag{AI.5}$$

Therefore, the average kernel based on Laplace transformation is equivalent to the population-averaged kernel.

**Appendix II: MATLAB Code**

Taking the shear coagulation kernel as example, the MATLAB code for the calculation of $Q_{n+1,n+1}$ and its error $\epsilon_n$ based on Gauss-Laguerre quadrature for the double integrals is listed as below:

```matlab
clear,
format long
n = 99;
for k = 1:n
    beta = zeros(k,k); [x,w] = Gauss_Laguerre(k); Q = 0;
    for i = 1:k
        for j = 1:k
            % collision kernel for shear coagulation
            beta(i,j) = ( x(i)^(1/3) + x(j)^(1/3) )^3;
            Q = Q + w(i)*w(j)*beta(i,j);
        end
    end
    p(k) = Q/2;
end
error = abs(p(2:n)-p(1:n-1));
figure, loglog(1:n,p,'.'),grid on
figure, loglog(1:n-1,error,'.'),grid on
function [x,w] = Gauss_Laguerre(n)
% n: Define the order of the Gauss-Laguerre quadrature
% x: the evaluation points of the nth Laguerre polynomial
% w: the weights of the nth Gauss-Laguerre quadrature
% w_i = x_i /((n+1)^2*(L_{n+1}(x_i))^2)
syms y;
p1 = laguerreL(n, y);
x = double(solve(p1 == 0, y));
w = zeros(n, 1);
for i = 1:n
    L_n_plus_1 = laguerreL(n+1, x(i));
    w(i) = double((x(i)/((n+1)^2*(L_n_plus_1)^2)));
end
end
```

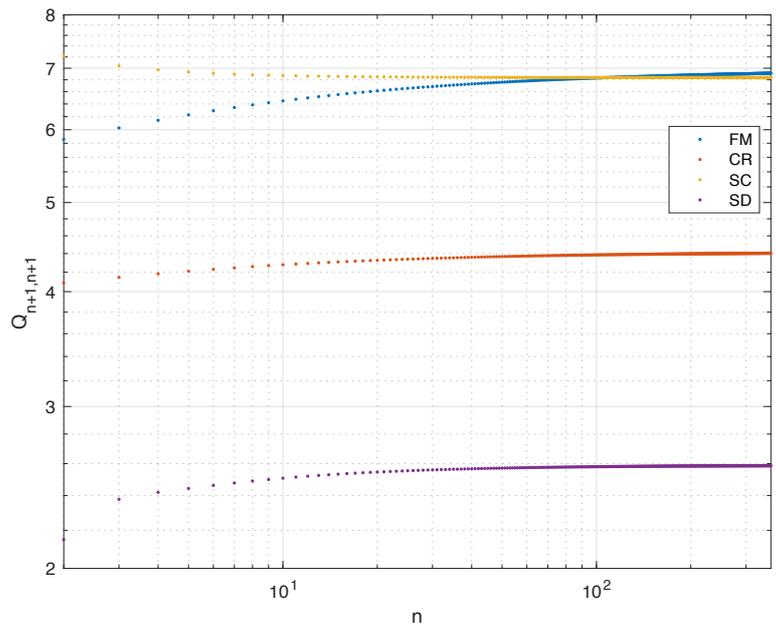

a

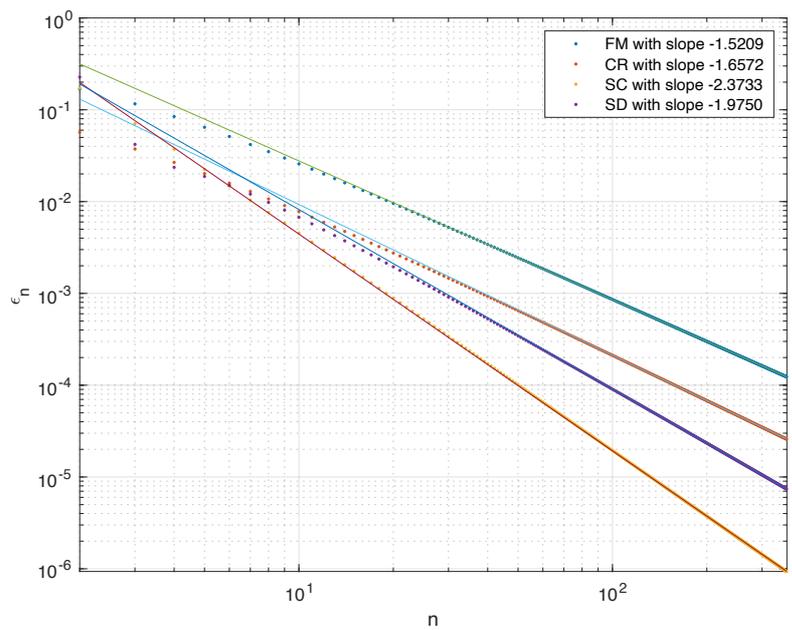

b

Fig.1. The numerical results of the Gauss Laguerre quadrature for the double integral. a) the change of the value $Q_{n+1,n+1}$ with $n$; b) the change of the *error* as the interpolation points increases.